\documentclass[12pt]{article}
\usepackage[right=1.25in,left=1.25in,top=1.1in,bottom=1.1in]{geometry}
\usepackage{hyperref}
\hypersetup{colorlinks, citecolor=blue, filecolor=blue, linkcolor=blue, urlcolor=blue}
\usepackage{graphicx}
\usepackage{url}
\usepackage[round]{natbib}
\usepackage{amsmath,amsthm} 
\usepackage{engord}
\usepackage{float}
\usepackage{pdflscape}
\usepackage{booktabs}
\usepackage{pgfplots}
\pgfplotsset{compat=1.14}
\pgfplotsset{every axis label/.append style={font=\tiny}}
\usepackage[labelsep=period]{caption} 

\usepackage{amssymb} 
\usepackage{multirow} 

\usepackage{xr}

\usepackage{setspace}
\usepackage{sectsty}
\sectionfont{\large}
\subsectionfont{\normalsize}
\subsubsectionfont{\normalsize}

\usepackage{graphicx}
\usepackage{subcaption}
\def\fillandplacepagenumber{%
 \par\pagestyle{empty}%
 \vbox to 0pt{\vss}\vfill
 \vbox to 0pt{\baselineskip0pt
   \hbox to\linewidth{\hss}%
   \baselineskip\footskip
   \hbox to\linewidth{%
     \hfil\thepage\hfil}\vss}}

\title{ \vspace*{-2.5cm} \hspace*{-0.5cm}Exchange Rate Pass-Through and Data Frequency: \\ Firm-Level Evidence from Bangladesh \footnote{
I am grateful to Dr. Georg Schaur for all his feedback on this study. 
}}

\author{Md Deluair Hossen\thanks{University of Tennessee, Knoxville.
\href{mailto:TK@TK.edu}{mhossen@utk.edu}} }

\date{ \vspace*{0.5cm} January, 2023\\
} 

\begin{document}

\bgroup
\let\footnoterule\relax

\begin{singlespace}
\maketitle

\begin{abstract}
    \noindent The vast literature on exchange rate fluctuations estimates the exchange rate pass-through (ERPT). Most ERPT studies consider annually aggregated data for developed or large developing countries for estimating ERPT. These estimates vary widely depending on the type of country, data coverage, and frequency. However, the ERPT estimation using firm-level high-frequency export data of a small developing country is rare. In this paper, I estimate the pricing to market and the exchange rate pass-through at a monthly, quarterly, and annual level of data frequency to deal with aggregation bias. Furthermore, I investigate how delivery time-based factors such as frequent shipments and faster transport affect a firm’s pricing-to-market behavior. Using transaction-level export data of Bangladesh from 2005 to 2013 and the Poisson Pseudo Maximum Likelihood (PPML) estimation method, I find very small pricing to the markets to the exchange rates in the exporter’s price. As pass-through shows how the exporters respond to macro shocks, for Bangladesh, this low export price response to the exchange rate changes indicates that currency devaluation might not have a significant effect on the exporter. The minimal price response and high pass-through contrast with the literature on incomplete pass-through at the annual level. By considering the characteristics of the firms, products, and destinations, I investigate the heterogeneity of the pass-through. The findings remain consistent with several robustness checks.\\


    
\end{abstract}
\end{singlespace}
\thispagestyle{empty}

\clearpage
\egroup
\setcounter{page}{1}


\section{Introduction\label{sec:introduction}}

\noindent The vast literature on exchange rate fluctuations estimates the exchange rate pass-through (ERPT). Most ERPT studies consider annually aggregated data for developed or large developing countries for estimating ERPT. These estimates vary widely depending on the type of country, data coverage, and data frequency \citep{vigfusson2009exchange,burstein2014international}. Several studies also document aggregation bias for ERPT estimates \citep{yoshida2010new,mumtaz2011exchange}. Furthermore, the ERPT estimation using firm-level high-frequency export data of a small developing country is rare. 

The ERPT literature focuses on various channels to estimate and explain incomplete ERPT, such as nominal price rigidity \citep{choudhri2006exchange}, currency invoice \citep{engel2006equivalence, gopinath2010currency, gopinath2010frequency}, productivity \citep{berman2012different, cook2014effect,li2015exchange}, quality \citep{chatterjee2013multi, chen2016quality}, variable markup \citep{knetter1989price,amiti2014importers, caselli2017multi} and distribution costs \citep{burstein2003distribution,corsetti2005macroeconomic,goldberg2010sensitivity}. However, most of these studies take away the goods' delivery time in international trade. The recent literature shows the importance of delivery time for understanding trade patterns \citep{berman2012time,li2019time}. This time difference between the slower and the faster but more costly delivery modes can create disconnect exchange rates. Considering delivery time differences at the importer end, \cite{aizenman2004endogenous} finds endogenous price to market (PTM) for exchange rate changes and provides predictions that faster and frequent shipments may explain the direction of incomplete ERPT.

In this paper, I answer two research questions. First, I investigate whether the ERPT estimates have aggregation bias. Following \cite{berman2012different} methods, I estimate the PTM at the exporter prices at monthly, quarterly, and annual frequency using transaction-level export data of Bangladesh from 2005 to 2013. This PTM estimates yield ERPT estimates. High-frequency data allows me to overcome the aggregation bias of ERPT estimates. Second, I investigate whether delivery time-based factors such as shipping frequency, per-shipment costs, distance, or fraction of air freight affect ERPT estimates. 

I find that the PTM estimate for the export prices is very low, 5\% at the monthly level, 3.5\% at the quarterly level and 4.2\% at the annual level. At the monthly level, for a 10\% real depreciation of BDT, the unit value price goes up by 0.47\%. This PTM estimates yield 95\%, 96.5\%, and 95.8\% ERPT estimates in the importer prices at monthly, quarterly, and annual levels, respectively. My ERPT estimates are at the upper bound of the literature with almost complete pass-through. The ERPT estimates show a very small aggregation bias depending on the data frequency. The pass-through increases for a longer horizon. This bias increases when I consider other firm characteristics. For example, I find that pass-through decreases more for firms with higher market share at a monthly level than at an annual level.

The pass-through increases if an exporter ships a product to a specific destination more frequently, ships more goods by air, and faces higher per-shipment costs in that destination country. In contrast, shorter distances and higher distribution costs decrease pass-through. Therefore, similar to the distribution costs channel for ERPT estimates in the literature, the delivery time-based characteristics are also crucial for the ERPT estimates.  

I also find that the exchange rates' price response is higher if the exporter is trading continuously for a longer horizon to a high-contract enforceability destination than the low enforceability destination. The sectoral analysis shows that capital, intermediate parts, semi-durable, and nondurable goods have a higher price response to the exchange rate changes than other sectors. Finally, I found mostly robust ERPT estimates for product and country-level data and different subsamples based on firm and destination characteristics in the robustness checks. 

These empirical findings provide several contributions to the existing literature.

First, the primary aggregation bias comes from the data frequency by averaging the prices of a firm's product for the entire year, ignoring high-frequency price changes and product churning. Another aggregation bias arises from the industry or country aggregation levels by ignoring firm characteristics and firm-level price response to the exchange rate changes. In this regard, \cite{dekle2007microeconomic} includes firm productivity and export share to find the exchange rate elasticity at both firm and aggregated levels using Japanese data. For data frequency, \cite{nakamura2012lost} provide a theoretical model to deal with ``product replacement bias" by emphasizing the importance of time horizon length for ERPT estimates. Extending this analysis, \cite{gagnon2014missing} uses transaction-level data at the annual level to overcome aggregation bias due to missing data on item turnover for constructing an aggregate price index. I estimate pass-through using similar transaction-level export data but at a smaller time horizon. I emphasize that ERPT estimates at various frequencies can shed light on the magnitude of aggregation biases.

Second, \cite{aizenman2004endogenous} shows that time-dependent transportation costs lead to the pre-buying of imports, which creates a disconnect between import prices and realized terms of trade. As the last-minute delivery mode (i.e., air shipments) is more expensive than the pre-buying delivery mode (i.e., ocean shipments), the imports are bought on the spot market only if the realized terms of trade are favorable enough. Transport mode and shipping frequency affect the likelihood of pre-buying and pricing to the market. Similar time-dependent transportation delinks export price and the exchange rate when the exporter sets the price based on the expected exchange rate during payment. For example, in the case of ocean shipment, the exporter may set the price based on the next month's expected exchange rate if he gets paid the next month after delivery. I show that high shipping frequency and fast transport lower pricing to the market using export price data. Thus, this study contributes to ERPT literature by including delivery time-based variables for ERPT estimations.

Third, higher per-shipment costs decrease shipping frequency \citep{kropf2014fixed, hornok2015administrative}. As the per-shipment fixed costs accrue through multiple shipments, the higher per-shipment costs increase non-sensitive cost components to exchange rate fluctuations and give heterogeneous ERPT estimates. In this regard, recent studies of ERPT focus on non-sensitive distribution cost to exchange rate fluctuations and use firm-level export data to explain heterogeneous ERPT \citep{berman2012different,chatterjee2013multi,li2015exchange}. Considering these costs, I find that the distribution cost's effect on the price response is smaller than the previous studies, potentially due to lower distribution costs for apparel sector exports.

Fourth, this study contributes to the pricing to market literature \citep{engel2006equivalence,atkeson2008pricing,goldberg2009micro, gopinath2010currency}. My empirical study is similar to \cite{fitzgerald2014pricing}, which focuses on goods produced in a common location and sells in multiple destinations to identify the extent of PTM. Considering the firm-level export price, my PTM estimates are remarkably lower than previous findings \citep{jaimovich2008product,fitzgerald2008exchange}. These low PTM estimates indicate that Bangladeshi exporters do not have the significant market power to change prices facing real exchange rate changes. Moreover, I provide PTM estimates for the entire economy accounting for firm-product-destination and year-fixed effects instead of focusing on a single industry \citep{goldberg2013structural,nakamura2010accounting}.

Fifth, for emerging economy ERPT estimates, most studies consider country-level annual data \citep{kabir1998exchange,hoque2004exchange,shambaugh2008new}. For example, by using aggregated annual data, \cite{bussiere2014exchange} finds the degree of ERPT to import and export prices for 40 countries, including 19 emerging market economies. Regarding aggregated high-frequency data, \cite{mallick2010data} uses two-digit industry-level aggregated data from India to find ERPT at monthly and annual levels. Considering disaggregated low-frequency data of China, \cite{li2015exchange} finds ERPT at the annual level. On this backdrop, my ERPT estimates for high-frequency disaggregated Bangladeshi exports show that some firm characteristics (e.g., fear of losing market share, dollar or euro-denominated currency price rigidity, frequency of price changes, and lack of short-run price negotiation power) may be more prominent in a small developing economy and may lead to a minimal price response to the exchange rate changes. 

Finally, understanding the pass-through is important to get key insights facing exchange rate changes, which allows us to evaluate the welfare of exporters, importers, and consumers. For example, the magnitude of pass-through can determine how currency devaluation affects exporter's pricing setting, consumer price, and inflation \citep{burstein2014international}. Furthermore, pass-through shows how the exporters respond to macro shocks. For Bangladesh, the export price response to the exchange rate changes is very low, indicating that currency devaluation might not significantly affect the exporter. 

The remainder of the paper is structured as follows. Section 2 describes firm-level trade data and country-level macro data. Section 3 shows the estimation strategy. Section 4 proceeds to the exchange rate pass-through estimation. Section 5 provides the robustness checks. Section 6 concludes. 

\section{Data\label{sec:data}}

This empirical exercise requires data for export prices, exchange rates, per-shipment costs, distribution costs, and some control variables. I get the export prices from transaction-level export data of Bangladesh. The bilateral exchange rates and export destination-specific data come from the International Monetary Fund (IMF) and World Bank. As I am interested in the ERPT estimates for various frequencies, I obtain these data at monthly, quarterly, and annual frequencies.

\subsection{Trade Data}
The data on transaction-level exports are from the customs authority of Bangladesh, the National Board of Revenue. The International Growth Center (IGC) collects and analyzes this dataset for policy analysis. They uploaded the dataset on their website for further studies.\footnote{The website link: https://www.theigc.org/country/bangladesh/data/.} From this dataset, I use the date of export, the exporter's unique identification number, the value of exports, the total volume of exports (in kilograms and units), eight-digit HS codes of the product, ports of export, and the destinations of the exports. To account for outliers, I trim the sample by winsorizing the export values smaller than one percentile and are higher than the 99 percentile at each data aggregation level.\footnote{In the robustness check, I include all observations.} For monthly-level data, I lose 30,859 observations, about 1.99\% of the total observations. Similarly, I lose 11,547 observations for yearly-level data, about 2\% of the total observations. 

Table \ref{ch3t1} shows the year-wise number of products, number of firms, export values, and number of observations. All export values are in constant 2010 Bangladeshi Currency, Taka (BDT). I convert these export values by using the GDP deflator to get concurrent values. Data shows considerable growth of export volume, except data anomaly of 2007.\footnote{There are some missing customs data in the dataset comparing the total export volume reported by the World Bank. I suspect that the sudden regime change of the government at that time might lead to this misreporting.} The number of firms grew moderately except in 2013. However, the number of products remains roughly the same, which indicates low product churning. The export share of the top ten destinations based on the export's total value is presented in table \ref{ch3t2}. The United States has the largest share of 23.49\%, followed by Germany's share of 15.69\% and the United Kingdom's share of 10.12\%. Almost 78\% of Bangladeshi exports are going to these top ten export destinations, all of which are members of the Organization for Economic Co-operation and Development (OECD). 

I do not observe the firm's export price data. I use per-unit value as a proxy for export price \citep{schott2004across,hallak2006product}. I can use export volume data in kilograms or units for per-unit value. Having a large share of missing data for the number of units, I use kilograms for calculating per-unit value in the baseline regressions. I calculate per-unit value, $UV_{ijk, m}$ from the export value, $V_{ijk, m}$ and export volume (in kilograms), $Q_{ijk, m}$. Thus, $UV_{ijk, m}=\frac{V_{ijk, m}}{Q_{ijk, m}}$. Here, the subscripts are defined as follows: $i$ is the firm; $j$ is the destination; $k$ is the HS eight-digits product; and $m$ is the month. The export value and volume are aggregated monthly from the daily transaction-level per-shipment value (free-on-board trade value) and weight. For quarterly data frequency, $UV_{ijk, q}=\frac{V_{ijk, q}}{Q_{ijk, q}}$, where $q$ is the subscript for the quarter. For annual data frequency, $UV_{ijk, t}=\frac{V_{ijk, t}}{Q_{ijk, t}}$, where $t$ is the subscript for year.\footnote{The number of unit-based per-unit values are considered in the robustness check. I use export volume in units, $N_{ijk, t}$, to calculate the per-unit value. Thus, $UV_{ijk, t}=\frac{V_{ijk, t}}{N_{ijk, t}}$.} Table \ref{ch3t3} shows summary statistics for the unit values, the number of products, and the number of destinations at the monthly and annual levels. The average monthly unit value is 20.35 USD per kilogram compared to the annual average unit value of 22.79 USD per kilogram. The average number of HS8 products per firm per destination per month is 2.15, almost half the annual value of 4.42. Similarly, the mean value of the number of destinations per firm per product per month is 2.56 compared to the annual value of 4.13. 

\subsection{Macro Data}
Monthly nominal exchange rates come from the IMF's International Financial Statistics (IFS). To find real exchange rate data at the monthly level, I calculate the following equation:
\begin{equation*}
    RER_{j,m} = NER_{j,m} \times\dfrac{CPI_{j,m}}{CPI_{BD,m}}
\end{equation*}

Where $NER_{j,m}$ is the bilateral nominal exchange rates of Bangladesh and country $j$ in month $m$, expressed in BDT for per-unit of $j$'s currency. $CPI_{j,m}$ and $CPI_{BD,m}$ are consumer price indices of country $j$ and Bangladesh in month $m$, respectively. An increase in $RER_{j,m}$ indicates a real depreciation of the Bangladeshi currency, BDT. For quarterly level, $ RER_{j,q} = NER_{j,q} \times CPI_{j,q}/CPI_{BD,q}$. For annual level, $RER_{j,t} = NER_{j,t} \times CPI_{j,t}/CPI_{BD,t}$. The CPIs are taken from the IFS database. 

Real GDP comes from the Center for International Prospective Studies (CEPII) Gravity dataset \citep{head2010erosion}. The bilateral distances come from the CEPII Geodist dataset \citep{mayer2011notes}. Distribution cost comes from \cite{goldberg2010sensitivity} and contains the distribution margin by destination and sector for a panel of 21 countries of OECD and 29 industries.

The per-shipment costs and the contract scores come from the DBI dataset of the World Bank's Doing Business Survey (DBS). DBS measures the ease of doing business in a country using the time and cost of documentary compliance, border compliance, and domestic transport for per-shipment costs, excluding tariffs. For enforcing the contract score, DBS measures the time and cost of resolving a commercial dispute.

\subsection{Exchange Rates of Bangladesh}

I plot the exchange rates against time to check how the exchange rates fluctuate through the study period. For a representative export destination, I choose the USA, the biggest export destination with 23.49\% export share. Thus, figures \ref{ch3f1} and \ref{ch3f2} show the monthly nominal and real exchange rates of Bangladeshi currency against USD, respectively. I find that for 2005-2007 and 2011-2013, the two curves are following similar trends. This co-movement of the nominal and real exchange rate is a stylized fact \citep{finn1999equilibrium}. However, for 2007-2011, the nominal and real exchange rates follow the opposite direction. While the nominal exchange rate is slightly increasing, the real exchange rate is falling significantly, which shows Bangladesh's deviation from purchasing power parity (PPP) with the USA. A potential reason for this deviation is the great recession of 2007-2008, where Bangladesh did not have any significant economic impact. During the recession and aftermath, the PPP deviated. In the robustness check, I split the sample into two parts based on the co-movement to capture the deviations from PPP. I found that the ERPT estimates are significantly different for these two periods. Figures \ref{ch3f3} and \ref{ch3f4} show annual nominal and real exchange rates of Bangladeshi currency against USD, respectively. Here, the deviation from purchasing power parity for Bangladesh and the USA is far smoother. Furthermore, figure \ref{ch3f4} shows considerable fluctuation for both real appreciation and depreciation of the exchange rate. I consider this fact in the robustness check to find asymmetric price responses for appreciation and depreciation.

\section{Estimation Strategy}

Following \cite{berman2012different}, I use the following three empirical models to find the exchange rate pass-through at monthly, quarterly, and annual levels:
\begin{equation} \label{ch3eqbasicr}
     \ln U V_{i j k, m}=\alpha+ \beta \ln R E R_{j, m} +D_{t}+D_{i j k} +\epsilon_{i j k, m}
\end{equation}
\begin{equation} \label{ch3eqbasicrq}
     \ln U V_{i j k, q}=\alpha+ \beta \ln R E R_{j, q} +D_{t}+D_{i j k} +\epsilon_{i j k, q}
\end{equation}
\begin{equation} \label{ch3eqbasicrt}
     \ln U V_{i j k, t}=\alpha+ \beta \ln R E R_{j, t} +D_{t}+D_{i j k} +\epsilon_{i j k, t}
\end{equation}

The dependent variable is $UV_{i j k, m}$, the unit price expressed in BDT for product $k$ of firm $i$ exporting to the country $j$ in a month $m$. $R E R_{j, m}$ is the average real exchange rate between Bangladesh and country $j$ in a month $m$. Using fixed effects properly, I can absorb heterogeneity based on firm productivity, goods elasticity, distribution costs, market size, trade costs, distribution channel, marketing quality, and unique taste. In all the models above, I consider year dummies $D_{t}$, which controls for time-varying macro-shocks common to all Bangladeshi exporters. Firm-product-destination fixed effects $D_{i j k}$ control for firm-product-destination based time-invariant unobservables. Finally, $\epsilon_{i j k, m}$ is the error term. Models \ref{ch3eqbasicrq} and \ref{ch3eqbasicrt} have similar notations, except $m$ is replaced by $q$, and $t$, quarter level and annual level data frequency, respectively.

The coefficient of interest is $\beta$. \cite{berman2012different} shows how the CES model with distribution costs paid in destination currency lead to a higher price response of a firm for real depreciation (see Appendix-\ref{c1}). I expect a positive $\beta$, which indicates that firms are pricing to the market, facing real depreciation, and firms set a higher price. From the pricing to the market, I can see how much price change happened for exchange rate changes at the exporter's end. The ERPT is calculated by $(1-\beta)$, which gives the ERPT of the import price in the destination's currency.\footnote{For example, if the BDT depreciates relative to USD and the BDT export price of a t-shirt is constant ($\beta=0$), then the import price of the t-shirt in USD drops. Therefore, the importer absorbs all the exchange rate changes.}

Now, I investigate how other factors regarding delivery time difference and distribution costs affect the PTM. I use the following specifications to find PTM at the monthly and annual levels:
    \begin{align} \label{ch3eqdistm}
         \ln U V_{i j k, m}=\alpha+\beta_{1} \ln R E R_{j, m}+\beta_{2} \ln R E R_{j, m} \times \ln X_{ijk, m}+D_{t}+D_{i j k} +\epsilon_{i j k, m}
    \end{align}
    \begin{align} \label{ch3eqdist}
         \ln U V_{i j k, t}=\alpha+\beta_{1} \ln R E R_{j, t}+\beta_{2} \ln R E R_{j, t} \times \ln X_{ijk, t}+D_{t}+D_{i j k} +\epsilon_{i j k, t}
    \end{align}
Where, $X_{ijk, m}$ or $X_{ijk, t}$ can be any of the following variables: shipping frequency for product $k$ of firm $i$ exporting to the country $j$ in a month $m$ or year $t$, per-shipment costs of the country $j$ in a year $t$, the distance of the country $j$, the fraction of air freight of firm $i$ exporting goods $k$ to the country $j$ in a month $m$ or year $t$, and distribution costs in the country $j$ for product $k$. The shipping frequency, per-shipment costs, distance, and fraction of air freight capture delivery time-based factors of the endogenous PTM model of \cite{aizenman2004endogenous} and affect ERPT. The distribution costs capture non-tradable local land and labor costs and affect ERPT \citep{burstein2003distribution}. In some specifications, I use market share, the number of products, and contract enforceability for $X_{ijk, t}$ to find heterogeneous ERPT estimates. The market share and the number of products capture productivity, which affects the ERPT. The contract enforceability captures the duration of relation-specific contracts, which affect pass-through \citep{bergin2001pricing}. The coefficient of interest is $\beta_2$. Thus, for faster transport or more frequent shipments, I expect $\beta_2<0$. On the other hand, for distribution costs, I expect $\beta_2>0$. Depending on the costs, the sign and magnitude of the coefficient will determine how these factors affect the ERPT estimates. 

In case of price rigidity, nominal exchange rate fluctuation can have a price response for the exporter \citep{gopinath2010currency}. To find PTM for the nominal exchange rate changes, I estimate the following model:
\begin{equation} \label{ch3eqbasicn}
     \ln U V_{i j k, m}=\alpha+ \beta \ln N E R_{j, m} +D_{t}+D_{i j k} +\epsilon_{i j k, m}
\end{equation}

Where $N E R_{j, m}$ is the nominal exchange rate between Bangladesh and country $j$ in a month $m$. Similarly, for quarterly and annual level specifications, all subscript $m$ will be replaced by subscript $q$ and $t$, respectively. With the interaction term, the model becomes
\begin{align} \label{ch3eqdistmn}
         \ln U V_{i j k, m}=\alpha+\beta_{1} \ln N E R_{j, m}+\beta_{2} \ln N E R_{j, m} \times \ln X_{ijk, m}+D_{t}+D_{i j k} +\epsilon_{i j k, m}
    \end{align}
    
For annual-level specifications, all subscript $m$ will be replaced by subscript $t$. Although the sign of the betas will be the same, here, the interpretation of betas changes for nominal exchange rates compared to betas of real exchange rate models. Here, PPP does not affect price response to the exchange rate change.

\section{Results}
In this section, I start with the benchmark regression results of PTM. Then I show how ERPT estimates vary depending on the characteristics of firms, products, and export destinations. Finally, I report results for robustness checks based on alternative specifications, subsamples, and different levels of aggregation. 

\subsection{Benchmark Regressions}
Table \ref{ch3tbasic} shows the result for the baseline regression of the estimation equations with the firm by product by destination and year fixed effects. The standard errors are robust and clustered with firm clusters to deal with any correlation among the observations within a specific exporter.\footnote{As the exchange rate is measured at the destination level, robust standard errors are also clustered with destination-by-year clusters. The paper does not report results as the significance level remains roughly similar.} In column 1, I report the main result for the real exchange rate changes at the monthly level data frequency. Here, I find that the price response is positive and significant. The per-unit value coefficient is 0.05, which is the PTM estimate. For a 10\% real depreciation of BDT, the unit value price goes up by $(1.1^{0.05}-1)\times 100=0.47\%$. Therefore, the ERPT at import price is 1-PTM \citep{berman2012different,chatterjee2013multi,chen2016quality,li2015exchange}. The ERPT estimate is 0.95, which means the exchange rate pass-through is 95\% at the importer price. By controlling demand effects with GDP, I find an even higher pass-through of 96\% in column 2. This nearly complete pass-through starkly contrasts the conventional incomplete ERPT estimates. Compared with the ERPT estimate of 46\% from \cite{campa2005exchange}, my ERPT estimate is much higher. However, higher pass-through can be explained by the fact that Bangladesh is an emerging economy with high financing costs and longer delivery times \citep{berman2012time} and have a large share of homogeneous goods in export \citep{aizenman2004endogenous}. My ERPT estimates are comparable with the almost complete ERPT findings of \cite{li2015exchange} for Chinese exports. Another explanation for such a minimal price response could be that the exporters do not have price negotiation power in a small developing country like Bangladesh.

At the quarterly level, columns 4 and 5 of table \ref{ch3tbasic} show positive, significant, and weaker price responses for real exchange rate changes than monthly level responses. Columns 7 and 8 show a little higher price response at the annual level of aggregation than the quarter level. For GDP control, the findings of the quarter level and annual level are roughly similar. For the quarter level, I find almost 97.5\% ERPT. This is in line with earlier findings of higher pass-through for longer horizon \citep{gopinath2010frequency,nakamura2012lost}. With more aggregated data, short-run pricing behavior is averaged out over a longer time horizon. However, at the annual level, ERPT increases by 0.8\% compared with the monthly level and decreases by 0.7\% compared with the quarterly level. This small ERPT difference from quarterly to annual data may come from aggregated exchange rates. In the context of the emerging economy, \cite{mallick2010data} finds a smaller ERPT (56\% monthly, 17.5\% annual) for low-frequency data of Indian exports and a very large aggregation bias. My aggregation bias is very small compared to this result.

Column 3 of table \ref{ch3tbasic} shows the per-unit value response to the monthly level's nominal exchange rate. I find a positive significant coefficient. For 10\% nominal depreciation, the unit value price goes up by $(1.1^{0.082}-1)\times 100=0.78\%$. This PTM is higher for the nominal exchange rate than the real exchange rate. I do not consider the price index for the nominal exchange rate, which can elevate price response. Surprisingly, Bangladeshi exporters are keeping their prices fixed, even facing nominal exchange rate depreciation. Therefore, the exchange rate shocks are absorbed into the import price. A potential explanation of nominal response is that wage in the labor-intensive garments industry (which covers more than 80\% export volume) is not inflation-adjusted and sticky.\footnote{In 1994, the monthly minimum wage for apparel workers was BDT 930, followed by BDT 1,662.50 in 2006, BDT 3,000 in 2010, and BDT 5,300 in 2013.} My ERPT estimate of 91.8\% is comparable to \cite{casas2020industry} ERPT estimate of 84.3\% for the nominal exchange rate. The nominal exchange rate's price response is weaker with more extended data frequency, as shown in columns 6 and 9. Thus, the pass-through increases from quarter to annual level aggregation, unlike real exchange rate changes. Aggregation bias is more prominent for nominal exchange rate changes than real exchange rate changes.

\subsection{Shipping Frequency and Air Transport}
Here, I consider how delivery time-based factors affect a firm's pricing behavior. I estimate the models \ref{ch3eqdistm}, \ref{ch3eqdist}, and \ref{ch3eqdistmn} for the per-shipment costs, shipping frequency, distances, fraction of air shipment, and distribution costs. Table \ref{ch3tdist} reports these results. Column 1 of table \ref{ch3tdist} finds a small significant negative effect of the shipping frequency on the price response at the monthly level. The effect is the same for the annual level data frequency, as reported in column 6. Thus, more frequent shipment decreases pricing to the market similarly at both monthly and annual levels. This is in line with the intuition that more frequent shipments increase the likelihood of buying at the spot market with the realized real exchange rate; hence the PTM decreases \citep{aizenman2004endogenous}. When I evaluate PTM at the minimum, mean, and maximum shipping frequency at the monthly level, PTM decreases by 0\%, 3.86\%, and 26.8\%, respectively. This shows that the effect of higher shipping frequency lowers the pricing to market to almost zero. The interaction term for the nominal exchange rate has the same coefficient in sign, significance, and magnitude, as reported in the bottom panel.

In column 2 of table \ref{ch3tdist}, I use the per-shipment costs and real exchange rate interaction and find a significant negative price response at the monthly level. Similarly, column 7 reports a larger response for the exchange rate and fixed-cost interactions at the annual level. Thus, higher fixed costs are reducing price response and increasing pass-through. This finding contradicts the intuition that the higher per-shipment costs reduce the shipping frequency \citep{hornok2015administrative,hornok2015per,hossen2020}, and increase price response. This can be explained by aggregation bias. The fixed cost price response comes from two mechanisms: distribution costs and aggregation bias. If the per-shipment costs (paid in foreign currency) capture non-sensitive costs components of the distribution costs, the higher per-shipment costs decrease pass-through. Column 2 shows that this mechanism is not working. Note that, at the monthly level, the export is almost at the transaction level; the fixed costs only capture the distribution costs aspect, not aggregation bias. Looking at column 7, I find that the aggregation mechanism is working at the annual level because shipping frequency has an equal effect on the price response at the monthly and annual levels. For the nominal exchange rate in the bottom panel, the interaction coefficient remains similar in sign, significance, and magnitude, comparing the findings for real exchange rate changes.

For distance interaction, I only consider observations with ocean shipments. I find significant negative effects at the monthly level in column 3 of table \ref{ch3tdist}. Column 8 reports a similar effect of distance at an annual level too. Thus, the price response is lower for the larger distance between the source and the destination. This finding contradicts the earlier findings regarding shipping frequency. As larger distance decreases shipping frequency \citep{hornok2015administrative}, the PTM should be higher. If I evaluate PTM at the minimum, mean, and maximum distance, the PTM coefficient reduces to 0.01, -0.002, and -0.01, respectively. PTM goes to negative territory at the furthest distance, potentially due to distance non-linearity. I find similar results for the nominal exchange rate, as shown in the bottom panel of the table \ref{ch3tdist}.

Air freight goods are time-sensitive \citep{hummels2013time} and more likely to be bought at the spot market, which lowers the PTM. I use the mode of transport to calculate the fraction of air freight goods for each exporter to a specific destination in a year. Column 4 shows that the exchange rate's price response decreases for a higher fraction of air freight goods. At an exporter's average air freight share (20.39\%), the PTM estimate is ($0.05-0.007\times 20.39)=-0.093$. Column 9 shows negative but non-significant air freight fraction interaction at the annual level.

Finally, for the distribution costs, columns 5 and 10 find a small positive effect of the distribution costs and exchange rate interactions on price. Note that the distribution margins from \cite{goldberg2010sensitivity} are only for 20 countries and limited sectors, which restricts the sample to only 73,006 observations at the monthly level and gives a strong price response. If I evaluate PTM at the minimum, mean, and maximum distribution cost, then PTM increases by 0.32\%, 0.59\%, and 0.64\%, respectively, at a monthly level. This is much smaller in magnitude compared to the distribution costs literature \citep{berman2012different,chatterjee2013multi,li2015exchange,chen2016quality}. The possible conclusion is that the distribution margin might not affect already minimal pricing to the exporter's market behavior. Furthermore, this small distribution cost effect on price may arise from a lower distribution margin of the apparel industry. I find a similar small interaction coefficient for the nominal exchange rate, as shown in column 1 of the bottom panel.

\subsection{Heterogeneous ERPT}
\subsubsection{Market share}
Lacking firm characteristics data, so far, I used firm fixed effects to account for heterogeneous marginal costs. Now, I consider market share and the number of products to capture the firm's mark-up adjustment \citep{atkeson2008pricing, amiti2014importers}. Furthermore, these two variables proxy for the firm's productivity \citep{tybout2003plant}. Market share represents the export ratio of an exporter to all firms' total export for an HS8 product to a specific export destination. The number of products captures HS8 products of a firm exported to a destination for a year. I run regressions with the firm's market share and the number of products to find the firm's heterogeneous price response to the exchange rate changes. The regression results are reported in table \ref{ch3tmarket}.

The interaction effects of the market share and the exchange rates are positive and significant for monthly and quarterly data, as reported in columns 1 and 2. Thus, the price response is higher for firms with a larger market share. The market share and real exchange rate interaction are weaker for quarterly data, as reported in column 2. This interaction becomes insignificant and negative for annual data, as reported in column 3. With more aggregated data frequency, the market share captures fewer pricing changes of the firm. However, the magnitude of the effect is very small. I find a positive, significant, and equal higher price response to the firms' real exchange rate with higher numbers of exported products for all data frequency levels. The export price increases by 0.3\% at average real exchange rate changes by doubling its average number of products, which shows very negligible pricing to market for larger product scope. Overall, I find evidence that larger, more productive firms can price to market more than small firms. 

\subsubsection{Contract enforceability}
Here, I examine a heterogeneous pass-through of the permanent firm based on the export destination's contract enforceability. A continuing exporter might want to keep the import price fixed for maintaining market share and find it easier to price to market for a higher contract enforceability destination \citep{bergin2001pricing} than a short-term exporter. \cite{campos2010incomplete} finds that the incumbent firms respond to the exchange rate changes differently than a new entrant. Therefore, I check whether the price response is higher for a firm that consistently exports a product for the entire study period. All columns of table \ref{ch3tcontract} consider permanent firms which export the same HS8 product to a specific country for the entire study period. Column 1 shows the ERPT for permanent firms and finds that the permanent firms' price response to the real exchange rate is much higher than the benchmark results.\footnote{As pricing setting behavior is driven by lagged exchange rates \citep{chen2016quality}, I investigate the effect of one year lagged real exchange rate on the exporter's price and do not find any significance for the lagged exchange rate effects, as reported in column 2.} Column 3 shows that the exchange rates coefficient increases by a small margin for a permanent firm exporting to a high contract score destination with a long-term contract. As the contract enforceability is captured with the contract score, a higher score means a high probability of enforcing the contract. Considering the magnitude of this interaction effect, the firms exporting to a high enforceable destination price to market marginally more to the exchange rate change than exporting to a weak enforceable destination. 

In column 4, I consider the exchange rate volatility of BDT against USD as previous studies predict very low PTM behavior for the highly volatile exchange rate \citep{aizenman2004endogenous,chatterjee2013multi}. I do not find any significant effect of the exchange rate volatility of BDT to USD. The statistically non-significant interaction coefficient results from a Bangladeshi exchange rate regime mainly targeted for improving the balance of trade with low volatility \citep{aziz2008role}. Moreover, \cite{corsetti2008high} suggests that nominal rigidities can create a low pass-through for high exchange rate volatility.

\subsubsection{ERPT estimates by sectors}

Demands and cost differences drive heterogeneous pass-through for different sectors or industries \citep{athukorala1994pricing}. Depending on the local distribution costs, the share of imported inputs, and the nature of the supply chain of the sectors, the exchange rate changes have different effects across the various product categories \citep{burstein2003distribution}. Table \ref{ch3tsector} presents the ERPT estimates for different sectors of exports. I categorize relevant sectors using broad economic categories \citep{hornok2015per}.\footnote{I use the following sectors: food and beverages (in short, Food), industrial supplies (Industrial), capital goods and transport equipment (Capital goods), parts and accessories thereof (Parts), durable, semi-durable, and nondurable goods.} Column 1 reports price responsiveness for exchange rate changes of seven sectors for monthly data frequency. I find a positive and significant response in all sectors except food sector goods, which negatively responds. The food sector price-setting behavior may be driven by monthly seasonality, which is not captured in year-fixed effects. I also find that parts (intermediate goods) have the highest price response, followed by capital goods and semi-durable goods. The price response is roughly similar for the annual level comparing the monthly level, as reported in column 2. I report the results of the HS2 industry analysis based on the top ten HS2 sectors in Appendix-\ref{c2}.

In summary, depending on the firm, product, and destination characteristics, the results of this section show statistical and economic significant heterogeneity in pass-through. However, magnitudes are overall small.

\section{Robustness Check}
Using the rich export data characteristics, I run four robustness checks at the monthly and annual levels. 

First, table \ref{ch3tsub} shows the robustness check based on the firm's characteristics. In column 1, I consider only single product firms (export one HS8 product in a year to a destination) and find a significant price response to the exchange rate changes. However, column 7 shows a non-significant price response at the annual level. Column 2 reports price response for the multi-product firms and shows a 94.7\% pass-through at the monthly level. This pass-through is smaller at the annual level, as reported in column 8. Columns 3 and 4 report the price response for the large firms with a total export value above the median and the small firms with a total export value below the median, respectively. I find that large firms have more pass-throughs than small firms. This shows that large firms have more ability to absorb the exchange rate changes in their markup. The annual level analysis shows a similar conclusion for firm size, as reported in columns 9 and 10. Columns 5 and 11 include all observations, including export values below one percentile and above the 99 percentile. The results do not change much from the benchmark results. Due to the deviations from the co-movement of the nominal and real exchange rates in 2007-2011, shown in figure \ref{ch3f4}, in columns 6 and 12, I only keep data from 2007 to 2011. Although I lose significance in determining ERPT at the monthly level, I find a strong price response at the annual level. The deviations from PPP and larger samples at the monthly level are driving noisy price responses. Moreover, the USD might not be a representative currency for choosing a sample splitting decision. 

Second, table \ref{ch3trobust} shows the robustness check based on the characteristics of the export destinations. Bangladesh's top ten export destinations are all OECD countries and responsible for 77.88\% of the total export share. I find a 93.4\% pass-through at the monthly level for all OECD countries, as reported in column 1. Column 7 shows a similar pass-through for OECD countries at the annual level data. Columns 2 and 8 consider only non-OECD countries and show a smaller significant price response at the monthly level. Still, the price response to the exchange rate changes is not significant at the annual level. This confirms the finding of \cite{li2015exchange} that export price is less responsive to non-OECD countries than OECD countries. Columns 3 and 9 show the price response for the countries which only use USD as their currency. I find a negative price response for only USD currency at the annual level. Columns 4 and 10 report insignificant price responses considering only Euro-denominated countries. Overall, a single currency cannot explain the price response.

I consider the pricing difference for real depreciation and real appreciation in columns 5 and 6 of table \ref{ch3trobust}. I keep observations for depreciation where the real exchange rate depreciated (an increase of $R E R_{j, t}$) compared to the last period. Column 5 shows a significant price response for real depreciation. I find a 93\% pass-through for real depreciation at the monthly level. For annual-level data, the price response for depreciation is smaller, as reported in column 11. Columns 6 and 12 show smaller price responses for real appreciation. One explanation could be for Bangladeshi currency; there are fewer occurrences and variations of appreciation than depreciation, as can be seen in figure \ref{ch3f2}. Thus, the asymmetric price responses for appreciation and depreciation suggest the firm's path-dependent pricing behavior \citep{thorbecke2010would,li2015exchange}.  

Third, table \ref{ch3tunits} shows the robustness check for alternative per-unit price calculation. Unit price calculation by weight is prevalent in the ERPT literature. However, some studies use the most frequently reported unit for every product-destination pair \citep{chatterjee2013multi}. I do not find consistent results for a price calculated with the units instead of kilograms at the monthly and annual levels, as reported in columns 2 and 4, respectively. Although the findings align with the literature for weight-based unit prices, the negative price response for unit-based analysis is puzzling. The bias due to one-third missing observations for unit-based per-unit value might lead to such price response. 

Finally, table \ref{ch3taggregate} reports the ERPT estimates at the different levels of aggregation. The firm-level, product-level, and country-level price responses to the real exchange rate changes at the monthly level are presented in columns 1, 2, and 3, respectively. Columns 1 and 2 show a positive significant price response. The product-level ERPT of 97.8\% is slightly larger than the firm-level ERPT of 95\%. As expected, the exchange rate change's price response does not have precision for the country-level aggregation at the monthly level. Comparatively larger coefficients are found at the annual level, as reported in columns 4-6. 

The main findings remain robust throughout the empirical exercises delineated above. I find a consistent and complete pass-through at the monthly and annual data frequency. 

\section{Conclusion}
This paper estimates the firm-level price response to the exchange rate at various levels of data frequency using transaction-level export data. The study finds very small pricing to the market at the monthly, quarterly, and annual levels. The pass-through at the importer price is 95\%, 96.5\%, and 95.8\% at the monthly, quarterly, and annual levels, respectively. Higher pass-thorough at the low-frequency data shows the presence of very small aggregation biases. The empirical analysis also finds that more frequent shipments and faster modes of transport increase pass-through. Therefore, delivery time-based factors are important to understand ERPT estimates. Furthermore, different characteristics of the firms and destinations yield heterogeneous and consistently high ERPT estimates. The findings are also robust to several data restrictions.   

These findings show that Bangladesh's exchange rate-based trade policies will have minimal effect on the exporter's pricing decisions. Overall, for a small developing country, nearly complete pass-through indicates that the exporters do not have significant market power for price-setting \citep{ghosh2007survey}. Although the magnitude is very low, my finding shows that the large exporters (in market share and product variety) can price to market more for exporting capital, semi-durables, and intermediate goods to a more contract-enforceable destination. Therefore, for a real depreciation shock, these exporters will gain more than the small firm exporting other category goods to less contract-enforceable destinations. 

The pass-through happens either by absorbing exchange rate changes in the markup or productivity enhancement by reducing marginal costs. Therefore, there may be issues of competitiveness, input costs, subsidy (ad valorem cash incentive on freight on board export value), and trade finance aid policies in Bangladesh, which may lower the price response to the exchange rate changes. For example, if the exporters receive cash incentives on the export value, they might show less price response to a real depreciation and keep the export price stable. For evaluating the impact of the exchange policies of a developing country, the firm's mark-up channel is an exciting direction for future research. Moreover, input costs or marginal cost heterogeneity should be considered for further ERPT studies.

\clearpage
\begin{singlespace}
\bibliographystyle{chicago}
\bibliography{main.bib}
\end{singlespace}

\newpage
\appendix
\setcounter{table}{0}
\renewcommand{\tablename}{Table}
\renewcommand{\figurename}{Figure}
\renewcommand{\thetable}{\arabic{table}}
\setcounter{figure}{0}
\renewcommand{\thefigure}{\arabic{figure}}

\section{Appendix Tables and Figures}
\begin{figure}[H]
    \centering
    \begin{subfigure}[t]{0.45\textwidth}
        \centering
        \includegraphics[width=70mm]{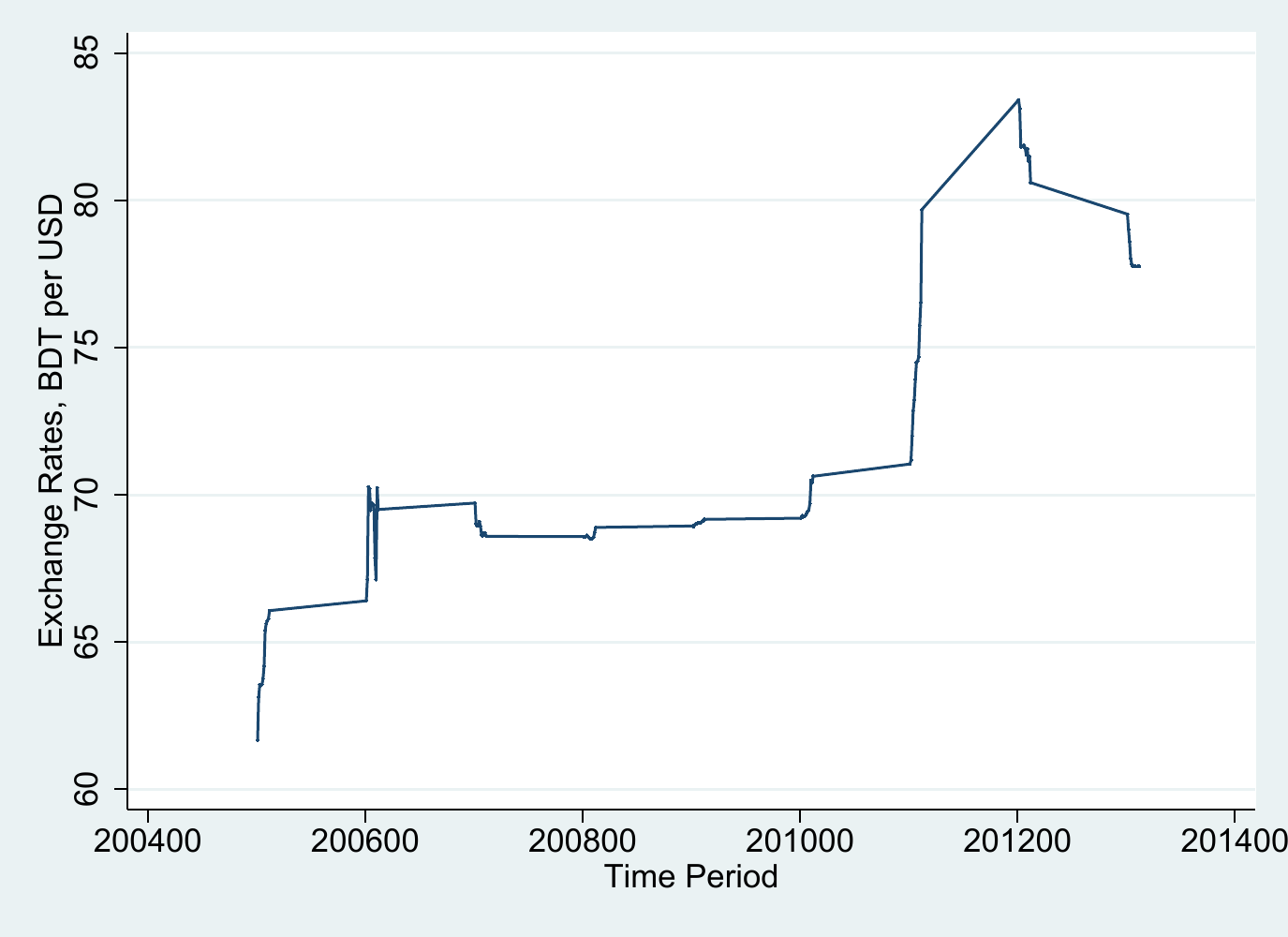}
        \caption{Monthly Nominal Exchange Rate of BDT against USD }
        \label{ch3f1}
    \end{subfigure}\hfill
        \begin{subfigure}[t]{0.45\textwidth}
        \centering
        \includegraphics[width=70mm]{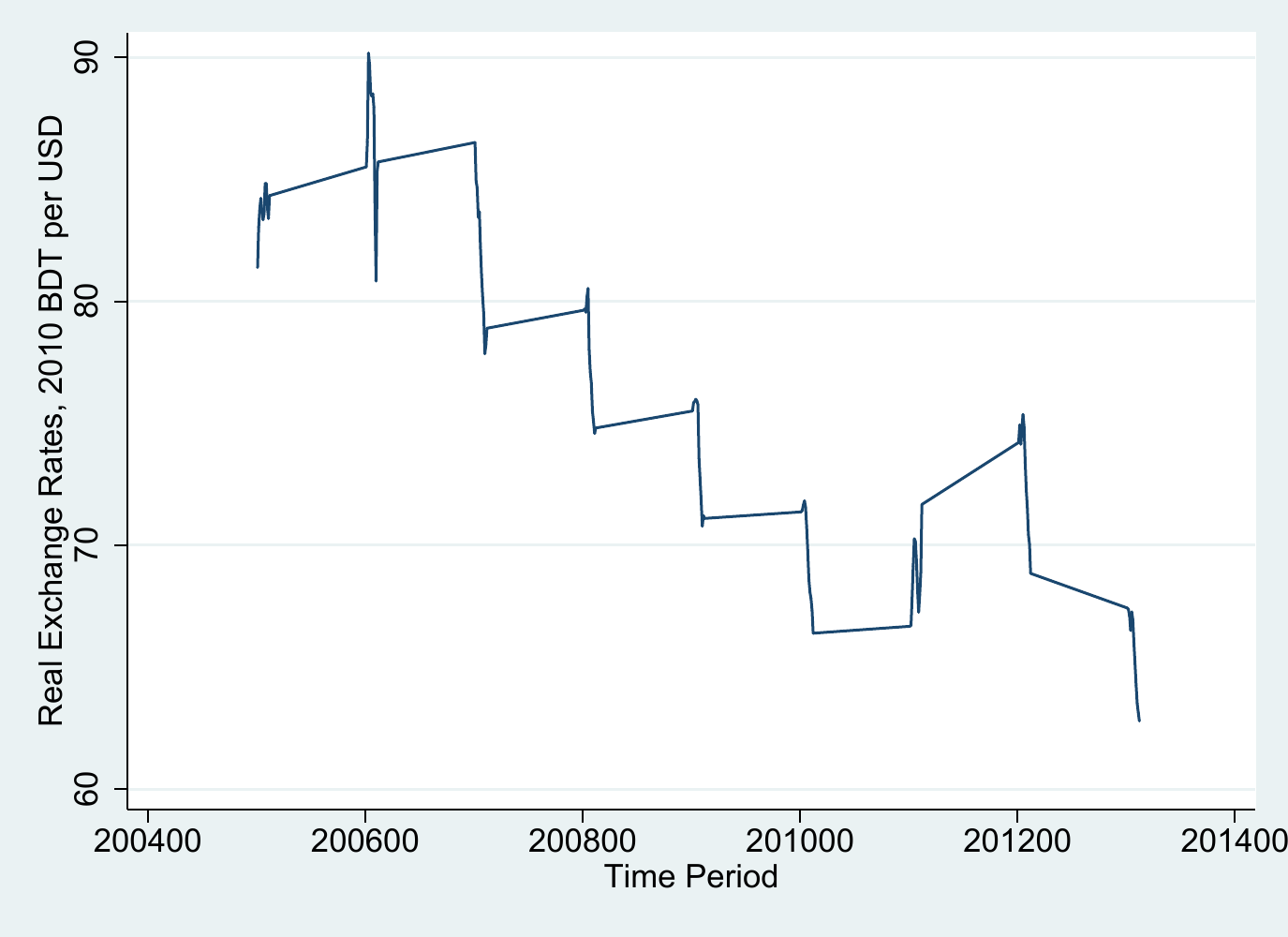}
        \caption{Monthly Real Exchange Rate of 2010 BDT against USD}
        \label{ch3f2}
    \end{subfigure}
    \medskip
    
    \begin{subfigure}[t]{0.45\textwidth}
        \centering
        \includegraphics[width=70mm]{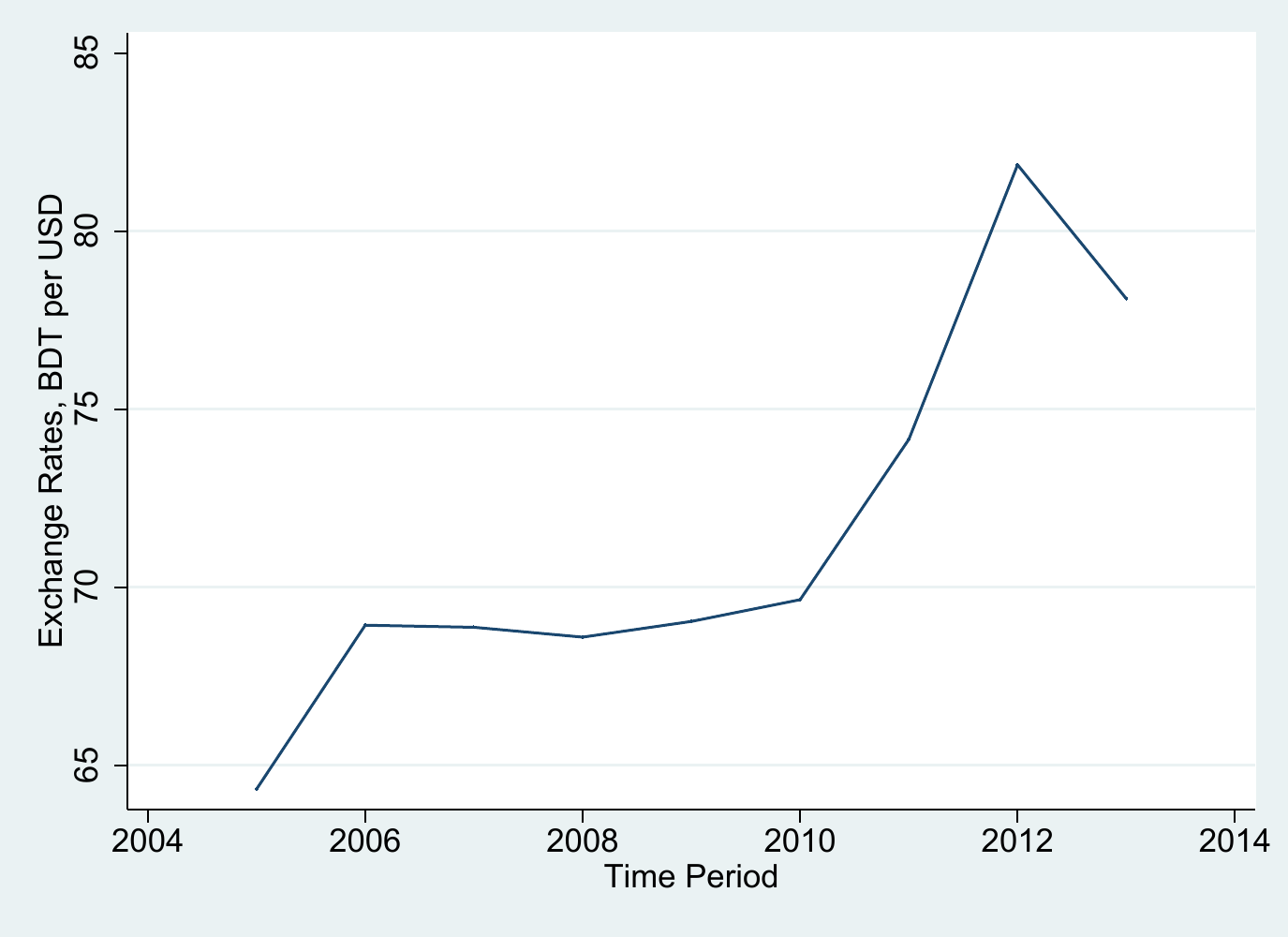}
        \caption{Annual Nominal Exchange Rate of BDT against USD}
        \label{ch3f3}
    \end{subfigure}\hfill
        \begin{subfigure}[t]{0.45\textwidth}
        \centering
        \includegraphics[width=70mm]{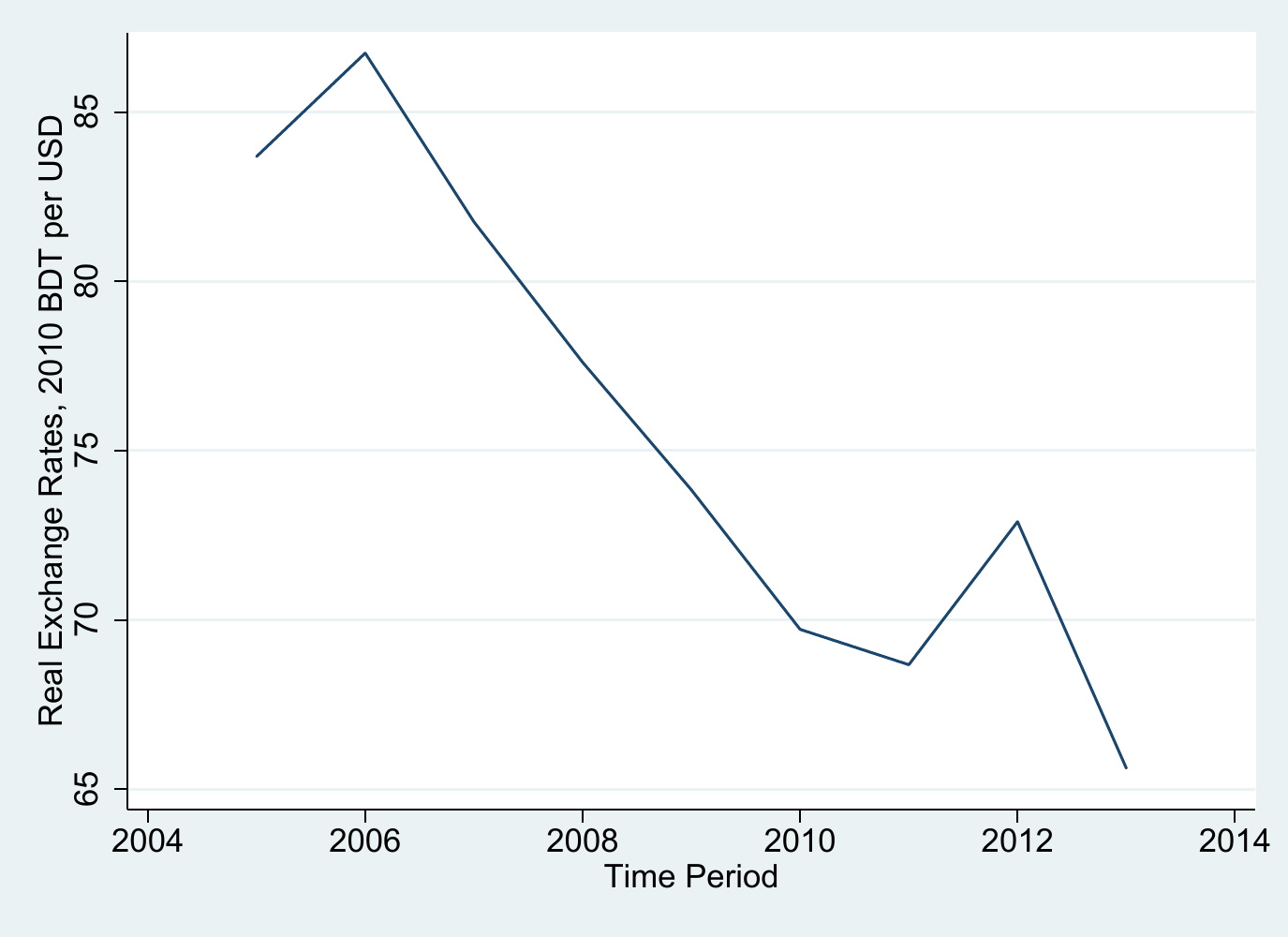}
        \caption{Annual Real Exchange Rate of 2010 BDT against USD}
        \label{ch3f4}
    \end{subfigure}
    
    \caption{Exchange Rate Fluctuations of BDT against USD}
    \label{ch3f11}
\end{figure}

\pagebreak
\begin{table}[H]
    
    \centering
    \singlespacing
    \scriptsize
    \caption{Exports Data by Year}
    \begin{tabular}{lcccc}
    \toprule
    Year & Firms & Products & Exports (USD) & Observations\\
    \midrule
         2005 & 5,235 &  1,658 & 8,055,902,197 & 330,431\\
          2006 & 5,568 & 1,779 & 11,689,611,679 & 406,727\\
           2007 & 5,781 &  1,611 & 8,394,929,235 & 301,404\\
            2008 & 6,528 & 1,703 & 14,685,512,111 & 482,781\\
             2009 &  6,637 & 1,721 & 14,455,759,090 & 504,451\\
              2010 & 6,880 & 1,868 & 18,501,543,521 & 623,735\\
               2011 & 7,163 & 1,718 & 24,837,277,321 & 655,419\\
                2012 & 7,550 & 1,826 & 28,042,176,843 & 764,868\\
                 2013 & 9,290 & 1,723 & 30,171,178,685 & 914,948\\
                 \midrule
                 Total & 18,910 & 4,351 & 158,833,885,184 & 4,984,764\\
                 \bottomrule \bottomrule
             
    \end{tabular}
    
    \label{ch3t1}
\end{table}

\begin{table}[H]
    \centering
    \scriptsize
    \singlespacing
    \caption{Top Ten Export Destinations of Bangladesh 2005-2013}
    \begin{tabular}{lc} \toprule
        Destination & Percentage of export (by value) \\
        \midrule
        United States & 23.49\\
        Germany & 15.69 \\
        United Kingdom & 10.12\\
         France & 6.34 \\
         Spain & 4.56 \\
          Canada & 4.22 \\
         Italy & 4.13\\
         Netherlands & 4.02 \\
         Belgium & 3.00\\
          Turkey &  2.44\\
        
         \bottomrule \bottomrule
\end{tabular} 
\label{ch3t2}
\end{table}
\begin{table}[H]
    \centering
    \singlespacing
    \scriptsize
    \caption{Summary Statistics}
    \begin{tabular}{lcccc|cccc}
    \toprule
     &  \multicolumn{4}{c|}{Monthly} & \multicolumn{4}{|c}{Annual} \\
    \cline{2-9}
      & Mean & Std. Dev. & Min & Max & Mean & Std. Dev. & Min & Max\\
    \midrule
    Unit values (USD
per kilograms)  & 20.35 & 783.09 & 0.0001 & 625653.4 & 22.79 & 1027.73  & 0.0003 & 569568.1 \\
      No. of products    & 2.15 & 1.91 & 1 & 48 & 4.42 & 5.14 & 1 & 72 \\
      No. of destinations    & 2.56 & 3.10 & 1 & 43 & 4.13  & 4.93 & 1 & 54\\
      \midrule
     Observations   &  \multicolumn{4}{c|}{1,512,048} & \multicolumn{4}{|c}{565,814} \\
            \bottomrule
            \bottomrule
            
    \end{tabular}
    
    \label{ch3t3}
\end{table}

\begin{landscape}

\begin{table}[H]
  \centering
  \scriptsize
  \singlespacing
  \caption{Response of the Export Prices to the Exchange Rates}
    \begin{tabular}{lccccccccc}
    \toprule
    & \multicolumn{3}{c|}{Monthly} & \multicolumn{3}{|c}{Quarterly} & \multicolumn{3}{|c}{Annual} \\
     \cmidrule(lr){2-4} \cmidrule(lr){5-7} \cmidrule(lr){8-10}
          & (1)   & (2)   & (3)   & (4)   & (5)   & (6)   & (7)   & (8)   & (9) \\
    
    \midrule
    
   Ln real exchange rate  & 0.050*** & 0.041*** &       & 0.035*** & 0.028** &       & 0.042*** & 0.033** &  \\
          & [0.010] & [0.010] &       & [0.010] & [0.011] &       & [0.014] & [0.015] &  \\
    Ln GDP  &       & 0.031* &       &       & 0.025 &       &       & 0.025 &  \\
          &       & [0.019] &       &       & [0.019] &       &       & [0.019] &  \\
    Ln nominal exchange rate   &       &       & 0.082*** &       &       & 0.064*** &       &       & 0.053*** \\
          &       &       & [0.012] &       &       & [0.011] &       &       & [0.013] \\
    Constant & 6.436*** & 5.588*** & 6.327*** & 6.470*** & 5.801*** & 6.369*** & 6.396*** & 5.732*** & 6.361*** \\
          & [0.035] & [0.513] & [0.041] & [0.035] & [0.520] & [0.040] & [0.048] & [0.521] & [0.045] \\
          
          \midrule
           Fixed Effects & \multicolumn{9}{c}{Firm-Product-Destination + Year}\\
         \midrule 
   
     Observations & 1,348,347 & 1,347,788 & 1,348,347 & 774,306 & 773,964 & 774,306 & 340,664 & 340,491 & 340,664 \\
    Adjusted R-squared & 0.897 & 0.897 & 0.897 & 0.896 & 0.896 & 0.896 & 0.891 & 0.891 & 0.891 \\
    \bottomrule

\bottomrule

    \end{tabular}%
  \label{ch3tbasic}%
\end{table}%

\vspace{-1em}
\footnotesize{Note: Separate OLS regressions of the estimation equations \ref{ch3eqbasicr}-\ref{ch3eqbasicrt} and \ref{ch3eqbasicn}. The dependent variable is the Ln per-unit value. The sample includes exports from Bangladesh to 164 countries in 4,351 HS8 products by 18,910 firms in 2005-2013. Standard errors are robust and clustered with firm clusters. ***$p < 0.01$; **$p < 0.05$; *$p < 0.1$.}
\fillandplacepagenumber
\end{landscape}
\begin{landscape}
\begin{table}[H]
  \centering
  \tiny
  \singlespacing
  \caption{Delivery Time-Based Price Response to Exchange Rate Changes}
    \begin{tabular}{lcccccccccc}
    \toprule
       & \multicolumn{5}{c|}{Monthly} &  \multicolumn{5}{|c}{Annual} \\
       \cmidrule(lr){2-6} \cmidrule(lr){7-11}
         & (1)   & (2)   & (3)   & (4)   & (5)   & (6) & (7) & (8)& (9) & (10)\\
    
    \midrule
    \textbf{Real Exchange Rate}    &       &       &       &       &       &       &       &  &&\\

    Ln real exchange rate  & 0.051*** & 0.151*** & 0.064*** & 0.050*** & 0.621*** & 0.038*** & 0.170*** & 0.065*** & 0.036*** & 1.114*** \\
          & [0.010] & [0.030] & [0.011] & [0.010] & [0.083] & [0.014] & [0.033] & [0.010] & [0.013] & [0.217] \\
    Ln real exchange rate $\times$ Ln shipping frequency & -0.003*** &       &       &       &       & -0.003*** &       &       &       &  \\
          & [0.000] &       &       &       &       & [0.000] &       &       &       &  \\
    Ln real exchange rate $\times$ Ln per-shipment costs &       & -0.011*** &       &       &       &       & -0.015*** &       &       &  \\
          &       & [0.004] &       &       &       &       & [0.004] &       &       &  \\
    Ln real exchange rate $\times$ Ln distance &       &       & -0.008*** &       &       &       &       & -0.008*** &       &  \\
          &       &       & [0.001] &       &       &       &       & [0.001] &       &  \\
    
    Ln real exchange rate $\times$ Air freight fraction &       &       &       & -0.007** &       &       &       &       & -0.006 &  \\
          &       &       &       & [0.003] &       &       &       &       & [0.004] &  \\
    Ln real exchange rate $\times$ Ln distribution costs &       &       &       &       & 0.002*** &       &       &       &       & 0.003*** \\
          &       &       &       &       & [0.000] &       &       &       &       & [0.001] \\
    Constant & 6.503*** & 6.426*** & 6.753*** & 6.478*** & 2.739*** & 6.492*** & 6.389*** & 6.705*** & 6.462*** & 0.413 \\
          & [0.034] & [0.045] & [0.004] & [0.034] & [0.385] & [0.048] & [0.061] & [0.003] & [0.047] & [0.988] \\
     \midrule
    Observations & 1,244,988 & 1,241,044 & 1,074,672 & 1,247,793 & 75,574 & 307,294 & 305,772 & 359,764 & 309,133 & 23,474 \\
    Adjusted R-squared & 0.896 & 0.896 & 0.851 & 0.896 & 0.658 & 0.890 & 0.890 & 0.841 & 0.891 & 0.622 \\
    \midrule
   \midrule
   
        \textbf{Nominal Exchange Rate}    &       &       &       &       &       &       &       &  &&\\
    
    Ln nominal exchange rate   & 0.080*** & 0.209*** & 0.063*** & 0.082*** & 0.756*** & 0.052*** & 0.202*** & 0.064*** & 0.049*** & 1.043*** \\
          & [0.011] & [0.032] & [0.010] & [0.011] & [0.084] & [0.013] & [0.035] & [0.010] & [0.012] & [0.200] \\
    Ln nominal exchange rate $\times$ Ln shipping frequency & -0.003*** &       &       &       &       & -0.003*** &       &       &       &  \\
          & [0.000] &       &       &       &       & [0.000] &       &       &       &  \\
    Ln nominal exchange rate $\times$ Ln per-shipment costs &       & -0.013*** &       &       &       &       & -0.016*** &       &       &  \\
          &       & [0.004] &       &       &       &       & [0.004] &       &       &  \\
    Ln nominal exchange rate $\times$ Ln distance &       &       & -0.008*** &       &       &       &       & -0.007*** &       &  \\
          &       &       & [0.001] &       &       &       &       & [0.001] &       &  \\
    
    Ln nominal exchange rate $\times$ Air freight fraction &       &       &       & -0.009** &       &       &       &       & -0.008** &  \\
          &       &       &       & [0.003] &       &       &       &       & [0.004] &  \\
    Ln nominal exchange rate $\times$ Ln distributions costs &       &       &       &       & 0.002*** &       &       &       &       & 0.003*** \\
          &       &       &       &       & [0.000] &       &       &       &       & [0.001] \\
    Constant & 6.401*** & 6.264*** & 6.752*** & 6.367*** & 2.141*** & 6.442*** & 6.302*** & 6.705*** & 6.417*** & 0.763 \\
          & [0.039] & [0.054] & [0.004] & [0.039] & [0.388] & [0.045] & [0.061] & [0.003] & [0.044] & [0.903] \\
    \midrule
    Observations & 1,244,988 & 1,241,044 & 1,074,672 & 1,247,793 & 75,574 & 307,294 & 305,772 & 359,764 & 309,133 & 23,474 \\
    Adjusted R-squared & 0.896 & 0.896 & 0.851 & 0.896 & 0.659 & 0.890 & 0.890 & 0.841 & 0.891 & 0.622 \\
    \midrule
    Firm by product by destination FE  &   Yes    &     Yes  &    No   &   Yes    &    No   &    Yes   & Yes      &  No & Yes & No\\
    Firm by destination FE  &    No   &   No    &   No    &   No    &   Yes    &    No   &    No   & No & No & Yes\\
    Firm by product FE  &    No   &   No    &   Yes    &   No    &   No    &    No   &    No   & Yes & No & No\\
    Year FE  &   Yes    &   Yes    &   Yes    &   Yes    &  Yes     &   Yes    &    Yes   &  Yes&Yes&Yes\\
    
    \bottomrule

    \bottomrule
    \end{tabular}%

  \label{ch3tdist}%
\end{table}%
\vspace{-1em}
\footnotesize{Note: Separate OLS regressions of the estimation equations \ref{ch3eqdistm}, \ref{ch3eqdist} and \ref{ch3eqdistmn}. The dependent variable is the Ln per-unit value. Standard errors are robust and clustered with firm clusters. ***$p < 0.01$; **$p < 0.05$; *$p < 0.1$.}
\fillandplacepagenumber
\end{landscape}
\begin{table}[H]
  \centering
  \singlespacing
  \scriptsize
  \caption{Market Share}
    \begin{tabular}{lccc}
    \toprule
          & Monthly   & Quarterly   & Annual \\
    & (1)   & (2)   & (3) \\
    \midrule

    Ln real exchange rate  & 0.042*** & 0.029*** & 0.042*** \\
          & [0.010] & [0.010] & [0.014] \\
    Ln real exchange rate $\times$ Market share & 0.008*** & 0.005*** & -0.002 \\
          & [0.001] & [0.001] & [0.001] \\
    Ln real exchange rate $\times$ Number of products & 0.003*** & 0.003*** & 0.002*** \\
          & [0.001] & [0.001] & [0.001] \\
    Constant & 6.435*** & 6.468*** & 6.393*** \\
          & [0.034] & [0.035] & [0.048] \\
    \midrule
    Observations & 1,348,347 & 774,306 & 340,664 \\
    Adjusted R-squared & 0.897 & 0.896 & 0.891 \\
    \bottomrule
    \bottomrule

    \end{tabular}%
  \label{ch3tmarket}%
\end{table}%

\vspace{-1em}
\footnotesize{Note: Separate OLS regressions of the estimation equations \ref{ch3eqdistm} and \ref{ch3eqdist}. The dependent variable is the Ln per-unit value. Firm by product by destination and year-fixed effects are used in all regressions. Standard errors are robust and clustered with firm clusters. ***$p < 0.01$; **$p < 0.05$; *$p < 0.1$.}
\begin{table}[H]
  \centering
  \scriptsize
  \singlespacing
  \caption{Contract Enforceability and Volatility}
    \begin{tabular}{lcccc}
    \toprule
          & (1)   & (2)   & (3)   & (4)    \\
     
    \midrule

  Ln real exchange rate  & 0.081*** & 0.040 & 0.003 & 0.049** \\
          & [0.029] & [0.036] & [0.046] & [0.025] \\
    Ln real exchange rate lagged &       & 0.058 &       &  \\
          &       & [0.043] &       &  \\
    Ln real exchange rate $\times$ Contract score &       &       & 0.001*** &  \\
          &       &       & [0.000] &  \\
    Ln real exchange rate $\times$ Volatility &       &       &       & 0.085 \\
          &       &       &       & [0.101] \\
    Constant & 6.375*** & 6.376*** & 6.426*** & 6.565*** \\
          & [0.111] & [0.145] & [0.147] & [0.096] \\
    \midrule
    Observations & 165,636 & 147,704 & 147,875 & 151,674 \\
    Adjusted R-squared & 0.858 & 0.855 & 0.856 & 0.861 \\
    \bottomrule

    \bottomrule
    \end{tabular}%
  \label{ch3tcontract}%
\end{table}%
\vspace{-1em}
\footnotesize{Note: Separate OLS regressions of the estimation equation \ref{ch3eqdist}. The dependent variable is the Ln unit value. Firm by product by destination and year-fixed effects are used in all regressions. Standard errors are robust and clustered with firm clusters. ***$p < 0.01$; **$p < 0.05$; *$p < 0.1$.}
\begin{table}[H]
  \centering
  \scriptsize
  \singlespacing
  \caption{Sectoral Analysis}
    \begin{tabular}{lcc}
    \toprule
     & \multicolumn{1}{c|}{Monthly} &  \multicolumn{1}{|c}{Annual} \\
      & (1)   & (2) \\
    \midrule

    Food & -0.051*** & -0.059*** \\
          & [0.013] & [0.017] \\
    Industrial goods & 0.037*** & 0.021 \\
          & [0.013] & [0.016] \\
    Capital goods & 0.075*** & 0.059** \\
          & [0.025] & [0.024] \\
    Parts & 0.111*** & 0.104*** \\
          & [0.033] & [0.026] \\
    Durables & 0.020 & 0.020 \\
          & [0.018] & [0.021] \\
    Semidurables & 0.068*** & 0.067*** \\
          & [0.011] & [0.015] \\
    Nondurables & 0.062*** & 0.060*** \\
          & [0.011] & [0.015] \\
    Constant & 6.411*** & 6.392*** \\
          & [0.040] & [0.054] \\
    \midrule
    Observations & 1,473,989 & 521,833 \\
    Adjusted R-squared & 0.851 & 0.818 \\
    \bottomrule

    \bottomrule
    \end{tabular}%
  \label{ch3tsector}%
\end{table}%
\vspace{-1em}
\footnotesize{Note: Separate OLS regressions of the estimation equations \ref{ch3eqbasicr}, and \ref{ch3eqbasicrt}. The dependent variable is the log of per-unit value. The independent variable is the log of the real exchange rate. Firm by product by destination and year-fixed effects are used in all regressions. Standard errors are robust and clustered with firm clusters. ***$p < 0.01$; **$p < 0.05$; *$p < 0.1$.}
\begin{landscape}
\begin{table}[H]
  \centering
  \tiny
  \caption{Robustness Check: Firm Characteristics }
    \begin{tabular}{lcccccccccccc}
    \toprule
    & \multicolumn{6}{c|}{Monthly} &  \multicolumn{6}{|c}{Annual} \\
     \cmidrule(lr){2-7} 
     \cmidrule(lr){8-13}
    &  Single     &  Multiple     &   Large &   Small  & Full-sample       & Subsample &  Single     &  Multiple     &   Large &   Small  & Full-sample      & Subsample \\
    &  product     &  product     &   firms    &  firms    &        &  &  product     &  product     &   firms    &  firms    &        &\\
    
         & (1)   & (2)   & (3)   & (4)   & (5)   & (6) & (7)   & (8)   & (9)   & (10)  & (11)  & (12) \\
    
    \midrule

   Ln real exchange rate  & 0.071** & 0.053*** & 0.041*** & 0.054*** & 0.049*** & 0.013 & 0.001 & 0.062*** & 0.039 & 0.064*** & 0.039** & 0.080*** \\
          & [0.035] & [0.011] & [0.014] & [0.011] & [0.011] & [0.013] & [0.021] & [0.021] & [0.029] & [0.016] & [0.017] & [0.021] \\
    Constant & 6.015*** & 6.436*** & 6.464*** & 6.442*** & 6.442*** & 6.445*** & 6.157*** & 6.428*** & 6.408*** & 6.350*** & 6.406*** & 6.174*** \\
          & [0.102] & [0.037] & [0.048] & [0.040] & [0.039] & [0.047] & [0.064] & [0.079] & [0.104] & [0.057] & [0.060] & [0.074] \\
    \midrule
    Observations & 16,833 & 1,310,547 & 610,086 & 642,828 & 1,375,823 & 702,308 & 62,117 & 247,549 & 106,373 & 157,162 & 348,662 & 175,030 \\
    Adjusted R-squared & 0.869 & 0.898 & 0.879 & 0.924 & 0.892 & 0.887 & 0.923 & 0.871 & 0.851 & 0.929 & 0.886 & 0.886 \\
    \bottomrule

    \bottomrule
    \end{tabular}%
  \label{ch3tsub}%
\end{table}%
\vspace{-1em}
\footnotesize{Note: Separate OLS regressions of the estimation equations \ref{ch3eqbasicr} and \ref{ch3eqbasicrt}. The dependent variable is the Ln per-unit value. Firm by product by destination and year-fixed effects are used in all regressions. Standard errors are robust and clustered with firm clusters. ***$p < 0.01$; **$p < 0.05$; *$p < 0.1$.}
\fillandplacepagenumber
\end{landscape}
\begin{landscape}
\begin{table}[H]
  \centering
  \tiny
  \caption{Robustness Check: Destination Characteristics}
    \begin{tabular}{lcccccccccccc}
    \toprule
     & \multicolumn{6}{c|}{Monthly} &  \multicolumn{6}{|c}{Annual} \\
      \cmidrule(lr){2-7} 
     \cmidrule(lr){8-13}
     &   OECD    &  Non-OECD     &   USD    &   Euro    &     Depreciation & Appreciation  &   OECD    &  Non-OECD     &   USD    &   Euro    &     Depreciation & Appreciation\\
     
          & (1)   & (2)   & (3)   & (4)   & (5)   & (6) & (7)   & (8)   & (9)   & (10)  & (11)  & (12) \\
    
    \midrule
          
   Ln real exchange rate  & 0.066*** & 0.030** & 0.102** & -0.000 & 0.071*** & 0.048*** & 0.068*** & 0.021 & -0.764*** & -0.002 & 0.049** & 0.017 \\
          & [0.014] & [0.013] & [0.050] & [0.009] & [0.017] & [0.010] & [0.020] & [0.019] & [0.235] & [0.018] & [0.023] & [0.020] \\
    Constant & 6.486*** & 6.106*** & 6.194*** & 6.740*** & 6.304*** & 6.489*** & 6.414*** & 6.049*** & 9.865*** & 6.698*** & 6.182*** & 6.618*** \\
          & [0.056] & [0.026] & [0.216] & [0.043] & [0.063] & [0.036] & [0.082] & [0.037] & [1.017] & [0.084] & [0.080] & [0.074] \\
     \midrule
    Observations & 1,031,948 & 316,399 & 172,240 & 464,747 & 582,506 & 683,241 & 260,648 & 80,016 & 43,073 & 121,853 & 135,047 & 126,568 \\
    Adjusted R-squared & 0.839 & 0.942 & 0.801 & 0.817 & 0.901 & 0.896 & 0.830 & 0.936 & 0.777 & 0.822 & 0.901 & 0.878 \\
    \bottomrule

    \bottomrule
    \end{tabular}%
  \label{ch3trobust}%
\end{table}%
\vspace{-1em}
\footnotesize{Note: Separate OLS regressions of the estimation equations \ref{ch3eqbasicr} and \ref{ch3eqbasicrt}. The dependent variable is the Ln per-unit value. Firm by product by destination and year-fixed effects are used in all regressions. Standard errors are robust and clustered with firm clusters. ***$p < 0.01$; **$p < 0.05$; *$p < 0.1$.}
\fillandplacepagenumber
\end{landscape}

\begin{table}[H]
  \centering
  \scriptsize
  \singlespacing
  \caption{Robustness Check: Kilograms vs Units}
    \begin{tabular}{lcccc}
    \toprule
    & \multicolumn{2}{c|}{Monthly} &  \multicolumn{2}{|c}{Annual} \\
     \cmidrule(lr){2-3} 
     \cmidrule(lr){4-5}
     & Kilograms     & Units   & Kilograms    & Units \\
          & (1)   & (2)   & (3)   & (4) \\
    
    \midrule
          
   Ln real exchange rate  & 0.050*** & -0.074*** & 0.042*** & 0.007 \\
          & [0.010] & [0.015] & [0.014] & [0.018] \\
    Constant & 6.436*** & 5.858*** & 6.396*** & 5.652*** \\
          & [0.035] & [0.058] & [0.048] & [0.070] \\
    \midrule
    Observations & 1,348,347 & 805,318 & 340,664 & 211,748 \\
    Adjusted R-squared & 0.897 & 0.800 & 0.891 & 0.723 \\
    \bottomrule

    \bottomrule
    \end{tabular}%
  \label{ch3tunits}%
\end{table}%
\vspace{-1em}
\footnotesize{Note: Separate OLS regressions of the estimation equations \ref{ch3eqbasicr} and \ref{ch3eqbasicrt}. The dependent variable is the Ln per-unit value. Firm by product by destination and year-fixed effects are used in all regressions. Standard errors are robust and clustered with firm clusters. ***$p < 0.01$; **$p < 0.05$; *$p < 0.1$.}
\begin{landscape}
\begin{table}[H]
  \centering
  \scriptsize
  \singlespacing
  \caption{Robustness Check: Different Levels of Aggregation}
    \begin{tabular}{lcccccc}
    \toprule
    & \multicolumn{3}{c|}{Monthly} &  \multicolumn{3}{|c}{Annual} \\
     \cmidrule(lr){2-4} 
     \cmidrule(lr){5-7}
     & Firm  & Product   & Country & Firm  & Product   & Country\\
          & (1)   & (2)   & (3)  & (4)   & (5)   & (6)\\
     
    \midrule

   Ln real exchange rate  & 0.050*** & 0.022** & -0.043 & 0.042*** & 0.032** & -0.055* \\
          & [0.010] & [0.010] & [0.028] & [0.014] & [0.015] & [0.029] \\
    Constant & 6.436*** & 6.341*** & 5.646*** & 6.396*** & 6.208*** & 5.457*** \\
          & [0.035] & [0.030] & [0.042] & [0.048] & [0.040] & [0.038] \\

    \midrule
    Firm by product by destination FE  &  Yes    &     No  &   No   &   Yes    &    No   &    No   \\
    Destination by product FE  &   No    &   Yes    &  No     &   No    &  Yes    &    No   \\
    Destination FE  &    No   &   No    &   Yes    &   No    &   No    &    Yes    \\
    Year FE  &   Yes    &   Yes    &   Yes    &   Yes    &  Yes     &   Yes    \\
     \midrule
    Firm SE clusters  &   Yes    &     No  &    No   &   Yes    &    No   &    No   \\
    Product SE clusters  &   No    &     Yes  &    No   &   No    &    Yes   &    No   \\
    Year SE clusters  &   No    &   No    &  Yes     &   No    &   No    &    Yes   \\
    \midrule
    Observations & 1,348,347 & 323,499 & 11,934 & 340,664 & 70,550 & 1,331 \\
    Adjusted R-squared & 0.897 & 0.840 & 0.616 & 0.891 & 0.802 & 0.676 \\
    \bottomrule

    \bottomrule
    \end{tabular}%
  \label{ch3taggregate}%
\end{table}%
\vspace{-1em}
\footnotesize{Note: Separate OLS regressions of the estimation equations \ref{ch3eqbasicr} and \ref{ch3eqbasicrt}. The dependent variable is the Ln per-unit value. ***$p < 0.01$; **$p < 0.05$; *$p < 0.1$.}
\fillandplacepagenumber
\end{landscape}

\newpage 
\section{Appendix One \label{c1}}
\renewcommand{\thetable}{B\arabic{table}}
\setcounter{table}{0}
\renewcommand{\thefigure}{B\arabic{figure}}
\setcounter{figure}{0}

\subsection{Basic ERPT Model}{c1}
\normalsize
Here, I reproduce the ERPT model of \cite{berman2012different} for different levels of aggregation effects. Consumption function for a representative agent in country $i$,
\begin{equation*}
    U=\Bigg[\int_{\Omega}^{} x_{i}(\varphi)^{\frac{\sigma-1}{\sigma}} d\varphi \Bigg]^{\frac{\sigma}{\sigma-1}}
\end{equation*}

Where, $x_{i}(\varphi)$ is the consumption of variety $\varphi$. $\Omega$ is the available varieties set and $\sigma$ is the elasticity. An exporter from a source country faces three transaction costs to export to the country $i$: an iceberg trade costs ($\tau_{i}$), fixed costs of exporting $F_{i}(\varphi)$ and distribution costs ($\eta_{i}$ units of labor paid at $\omega_{i}$ wage rate in destination currency). Then, the consumer price in country $i$ is, $p_{i}^{c}\equiv \frac{p_{i} \tau_{i}}{\epsilon_{i}}+\eta_{i}\omega_{i}$. $p_{i}$ is the exporter currency price, and $\epsilon_{i}$ is the nominal exchange rate between the exporter's country and country $i$. The quantity demanded is 
\begin{equation*}
    x_{i}(\varphi)=Y_{i}P_{i}^{\varphi-1} [p_{i}^{c}]^{-\sigma}
\end{equation*}
 
Where $Y_{i}$ and $P_{i}$ are the income and price level of country $i$. The cost of the exporter for exporting $ x_{i}(\varphi) \tau_{i}$ is
 \begin{equation*}
     \frac{\omega x_{i}(\varphi) \tau_{i}}{\varphi}+F_{i}(\varphi)
 \end{equation*}
 
 Therefore, the profit-maximizing price for the exporter is
 \begin{equation*}
     p_{i}(\varphi)=\frac{\sigma}{\sigma-1} \big(1+\frac{\eta_{i}q_{i}\varphi}{\sigma \tau_{i}}\big) \frac{\omega}{\varphi} 
 \end{equation*}
 
 Where, real exchange rate, $q_i\equiv \frac{\epsilon_i \omega_i}{\omega}$. Thus, the price elasticity
 \begin{equation} \label{ch3eqap1} \tag{3.9}
     e_{p_{i}}(\varphi)=\dfrac{dp_{i}(\varphi)}{dq_{i}}\dfrac{q_{i}}{p_{i}(\varphi)}=\dfrac{\eta_{i}q_{i}\varphi}{\sigma \tau_{i}+\eta_{i}q_{i}\varphi}
 \end{equation}
 
By controlling $\varphi$, $\eta_{i}$, $\tau_{i}$ and $\sigma$, I find that $e_{p_{i}}(\varphi)$ only depends on the $q_{i}$. Thus, for the price response to the exchange rate, the estimation equation \ref{ch3eqbasicr} comes from equation \ref{ch3eqap1}. The frequency of data only affects the real exchange rate. Depending on the monthly, quarterly, or annual level data, I get a different estimate for $e_{p_{i}}(\varphi)$.

\newpage
\section{Appendix Two
\label{sec:appendix:two}}
\renewcommand{\thetable}{C\arabic{table}}
\setcounter{table}{0}
\renewcommand{\thefigure}{C\arabic{figure}}
\setcounter{figure}{0}

\subsection{Industry Analysis\label{c2}}

I find a positive significant price response to the exchange rates change for the knitted and non-knitted apparel, textile, footwear, and mineral fuel industries, as reported in table \ref{ch3tindustry}. As the non-knitted apparel industry covers 40\% of the export share, this sector is driving the price response. As the non-knitted garment sector is a little higher up in the value chain than the knitted sectors (T-shirts), the price response to the exchange rate changes is more than two times higher. Mineral fuel shows the highest price response, followed by textile and non-knitted apparel sectors facing real exchange changes.

\begin{landscape}
\begin{table}[H]
  \centering
  \singlespacing
  \scriptsize
  \caption{Industry Analysis}
    \begin{tabular}{llc|c}
    \toprule
         
     HS2 & Industry & Export Share & Ln Unit Value \\
    \midrule
         
    61 & Articles of apparel and clothing accessories, knitted or crocheted & 0.405  & 0.046* \\
        &&  & [0.025] \\
    62 & Articles of apparel and clothing accessories, not knitted or crocheted & 0.400  & 0.102*** \\
        &&  & [0.029] \\
  
    63 & Other made up textile articles & 0.037  & 0.182*** \\
        &&  & [0.052] \\
    53 & Other vegetable textile fibers & 0.030  & 0.049 \\
        &&  & [0.034] \\
    3 & Fish and crustaceans, mollusks and other aquatic invertebrates & 0.025  & 0.216 \\
        &&  & [0.138] \\
    41 & Raw hides and skins (other than fur skins) and leather & 0.014  & 0.087 \\
        &&  & [0.055] \\
    64 & Footwear, gaiters and the like; parts of such article & 0.013  & 0.074** \\
        &&  & [0.036] \\
  
    27 & Mineral fuels, mineral oils, and products of their distillation & 0.006  & 0.214*** \\
        &&  & [0.019] \\
    87 & Vehicles other than railway or tramway rolling-stock, and parts  & 0.006  & 0.310 \\
        &&  & [0.462] \\
    58 & Special woven fabrics; tufted textile fabrics; lace & 0.005  & 0.087 \\
        &&  & [0.231] \\
    
    \bottomrule
    \bottomrule
    \end{tabular}%
  \label{ch3tindustry}%
\end{table}%
\vspace{-1em}
\footnotesize{Note: Separate OLS regressions of the estimation equation \ref{ch3eqbasicrt}. The dependent variable is the Ln unit value. The independent variable is the Ln real exchange rate. Firm by product by destination and year-fixed effects are used in all regressions. Standard errors are robust and clustered with exporter clusters. ***$p < 0.01$; **$p < 0.05$; *$p < 0.1$.}
\fillandplacepagenumber
\end{landscape}

\end{document}